\documentclass[
aps,prd,
12pt,%10pt
nopreprintnumbers,
showpacs,
eqsecnum,
nofootinbib
]{revtex4-1}

\usepackage{graphicx}
\usepackage{amssymb}

\begin{document}

\title{Analytical Approximation for Newtonian Boson Stars in Four and Five
Dimensions\\ {\small ---A Poor Person's Approach to Rotating Boson
Stars---}}
\author{Nahomi Kan}\email[]{kan@gifu-nct.ac.jp}
\affiliation{National Institute of Technology, Gifu College,
Motosu-shi, Gifu 501-0495, Japan
}
\author{Kiyoshi Shiraishi}\email[]{shiraish@yamaguchi-u.ac.jp}
\affiliation{
Graduate School of Sciences and Technology for Innovation, Yamaguchi
University, Yamaguchi-shi, Yamaguchi 753--8512, Japan}
\date{\today}
%\date{}

\begin{abstract}
In this paper, we study rotating boson stars in the large coupling limit
as well as in the Newtonian limit. 
We investigate
the equilibrium solutions 
%in the weak gravitational limit 
in
four and five dimensions by adopting some analytical approximations.
We show that the relations among the radius, angular momentum, 
Newtonian energy, and quadrupole moment (for the four-dimensional
solution) of the boson star can be qualitatively realized for the minimal
number of boson-star parameters.
\end{abstract}

%\preprint{}


\maketitle

%%%%%%%%%%%%%%%%%%%%%%%%%%%%%%%%%%%%%%%%%%%%%%%%%%%%%%%%%%%%%%%%%%%%%%%%%%%
%Introduction
%%%%%%%%%%%%%%%%%%%%%%%%%%%%%%%%%%%%%%%%%%%%%%%%%%%%%%%%%%%%%%%%%%%%%%%%%%%
%%%%%%%%%%%%%%%%%%%%%%%%%%%%%%%%%%%%%%%%%%%%%%%%%%%%%%%%%%%%%%%%%%%%%%%%%%%
\section{Introduction}
\label{sec1}
%%%%%%%%%%%%%%%%%%%%%%%%%%%%%%%%%%%%%%%%%%%%%%%%%%%%%%%%%%%%%%%%%%%%%%%%%%%
%%%%%%%%%%%%%%%%%%%%%%%%%%%%%%%%%%%%%%%%%%%%%%%%%%%%%%%%%%%%%%%%%%%%%%%%%%%

Along with the development of observational cosmology, including the
recent gravity-wave detection, the theoretical study of compact objects
is actively advancing apace.  
Thus far, remarkable progress
has been made in the study of universal relations between physical
quantities of compact objects, such as 
the angular momentum, 
quadrupole moment, mass, and radius. 

The study of self-gravitating systems is very interesting 
because the characteristics of general relativity or the modified
gravity appear in the physical quantities of the system. 
For example, Brihaye, Hartmann, and collaborators argued about a
boson star solution in a higher-dimensional spacetime \cite{BH,HRS,HKKL}.

From the perspective of particle cosmology, the boson star
\cite{Jetzer,LM,SM} is one of the
 candidates for dark matter \cite{R1SL1,R1SL2,TCL,ST,MA}. 
The ($1+3$)-dimensional boson
star was studied as the simplest model of a self-gravitating system and
the Newtonian treatment of gravitating bosons has been often discussed
\cite{Jetzer}.  
Although the difference between the Newtonian treatment and the general
relativistic model is significant, the Newtonian treatment is known to 
enable the evaluation of the mass of the boson star qualitatively. 
It is necessary to
examine the Newtonian approximation in higher-dimensional boson
systems, since a substantial understanding is important to 
check whether a universal property  of compact objects exists
even in the modified gravity theory. 

We discuss the qualitative behavior of models such as a
four-dimensional rotating boson star (which has been studied in the
general relativistic model
\cite{R1S,R1SM1,R1SM2,Ryan,HRR}), a boson star in dimensions with
compactified space and five-dimensional boson stars using Newtonian
approximation.  We study the system of a scalar field with large
self-interaction
\cite{colpi} in this paper. A deep understanding of the basic aspects of
self-gravitating systems, which is independent of a possible
correction in the gravity theory, is expected from this study.

The present paper is organized as follows.
In Sec.~\ref{sNL}, we obtain the action, the Hamiltonian, and field
equations for a model of a self-interacting, gravitating boson field in
the Newtonian limit.
The large coupling limit of the model is defined in Sec.~\ref{sec3} and
we obtain the spherical solution for a boson star and discuss its mass. 
In Sec.~\ref{sec4}, approximated solutions for rotating boson stars are
obtained, and the relation among their physical quantities is
studied. The boson star in the Kaluza-Klein background is considered in
Sec.~\ref{sec5}, and the stability against a small variation along with
an extra dimension is discussed. In Sec.~\ref{sec6}, five-dimensional
boson stars in the Newtonian limit are investigated.
The last section is devoted to a summary and future prospects.

%%%%%%%%%%%%%%%%%%%%%%%%%%%%%%%%%%%%%%%%%%%%%%%%
%%%%%%%%%%%%%%%%%%%%%%%%%%%%%%%%%%%%%%%%%%%%%%%%
\section{The Newtonian limit}
\label{sNL}
%%%%%%%%%%%%%%%%%%%%%%%%%%%%%%%%%%%%%%%%%%%%%%%%
%%%%%%%%%%%%%%%%%%%%%%%%%%%%%%%%%%%%%%%%%%%%%%%%

We consider a system of self-interacting, gravitating scalar bosons
of mass $m$ governed by the following field-theoretical action; 
\begin{equation}
S=\int d^4x\,{\sqrt{-g}}\,{\cal L}=\int
d^4x{\sqrt{-g}}\left[\frac{1}{16\pi G}R- |\partial\phi|^2-m^2|\phi|^2-
\frac{1}{2}\tilde{\lambda}|\phi|^4\right]\,,
\end{equation}
where $d^4x=dt\,d^3\mbox{\boldmath $x$}$ , $G$ is the Newton constant,
$R$ is the Ricci scalar,
$\tilde{\lambda}$ is the dimensionless scalar self-coupling constant, and
$|\partial\phi|^2\equiv
g^{\mu\nu}(\partial_\mu\phi)^*(\partial_\nu\phi)$.

By the variational principle, we derive the Einstein equation from the
action as
\begin{equation}
R^\mu_\nu-\frac{1}{2}\delta^\mu_\nu R=8\pi G T^\mu_\nu\,,
\end{equation}
where the energy-momentum tensor in the system is given by
\begin{equation}
T_{\mu\nu}=\Big[\partial_\mu\phi^*\partial_\nu\phi+
\partial_\nu\phi^*\partial_\mu\phi-
g_{\mu\nu}|\partial\phi|^2-g_{\mu\nu}
m^2|\phi|^2\Big]
-g_{\mu\nu}
\frac{1}{2}\tilde{\lambda}|\phi|^4\,.
\end{equation}
The equation of motion for the complex scalar field $\phi$ is given by
\begin{equation}
\Box\phi-m^2\phi-\tilde{\lambda}|\phi|^2\phi=0\,,
\label{basiceq}
\end{equation}
where $\Box\phi\equiv\frac{1}{\sqrt{-g}}\partial_\mu
(\sqrt{-g}g^{\mu\nu}\partial_\nu
\phi)$ is the covariant d'Alembertian.

The Newtonian limit can be attained by
assuming that the spacetime metric in the weak field approximation
can be written as
\begin{equation}
g_{00}\approx -\left(1+%\frac
{2\Phi}%{c^2}
\right)\,,\quad \sqrt{-g}\approx
1\,,
\end{equation}
where $\Phi$ is the Newtonian gravitational potential.

Assuming further that a complex scalar field has a nearly harmonic time
dependence expressed by%
\footnote{Since we consider the Newtonian limit, the frequency $\omega$
is close to the scalar field mass $m$.} %R1
\begin{equation}
\phi=\frac{1}{\sqrt{2m}}\psi(\mbox{\boldmath $r$}, t)\,e^{-imt}\,,
\end{equation}
we obtain the (non-linear) Schr\"odinger equation
\begin{equation}
i\dot{\psi}=-\frac{1}{2m}\nabla^2\psi+m\Phi\psi+
\frac{\tilde{\lambda}}{4m^2}|\psi|^2\psi\,
\end{equation}
as the Newtonian limit of Eq.~(\ref{basiceq}), where $\nabla^2$ is the
Laplacian in the flat space and the dot indicates the time derivative.
In the present limit, the Einstein equations reduce to the Poisson
equation
\begin{equation}
\nabla^2\Phi=4\pi Gm|\psi|^2\,.
\end{equation}

The Newtonian treatment of the Lagrangian and Hamiltonian is as follows.
We find the following Newtonian action in the limit:
\begin{equation}
S\cong\int dt\, d^3\mbox{\boldmath $r$}
\left[-\frac{1}{8\pi G}({\nabla}
\Phi)^2+i\psi^*\dot{\psi}-
\frac{1}{2m}|\nabla\psi|^2-m\Phi |\psi|^2-
\frac{1}{8}\frac{\tilde{\lambda}}{m^2}|\psi|^4
\right]\,,
\end{equation}
where $(\nabla\Phi)^2\equiv {\bf\nabla}{\Phi}\cdot{\bf\nabla}{\Phi}$
and the symbol $\cong$ indicates that some surface terms have been
omitted.
Therefore, the Hamiltonian of the system is derived as
\begin{equation}
\hat{H}(\psi,\Phi)=\int d^3\mbox{\boldmath $r$}\,{\cal H}
=\int
d^3\mbox{\boldmath
$r$}
\left[\frac{1}{8\pi G}({\nabla}
\Phi)^2+
\frac{1}{2m}|\nabla\psi|^2+m\Phi |\psi|^2+
\frac{1}{8}\frac{\tilde{\lambda}}{m^2}|\psi|^4
\right]\,.
\label{NE}
\end{equation}

On the other hand, the Newtonian number of particles
is expressed as
\begin{equation}
\hat{N}\equiv\int d^3\mbox{\boldmath $r$}\,|\psi|^2\,.
\label{ndef}
\end{equation}

In addition, we require the condition $\hat{N}=N$,
i.e., the condition that the system contains $N$ scalar bosons.
Then, we consider
${\delta}\{\hat{H}-\mu(\hat{N}-N)\}=0$ as an equation
for the scalar matter field in the mean field approximation, where $\mu$
is a Lagrange multiplier.

Now, we obtain two coupled equations for the stationary gravitational
field and the matter field as follows:
\begin{eqnarray}
& &\nabla^2\Phi=4\pi G m|\psi|^2\,,
 \\
& &-\frac{1}{2m}\nabla^2\psi+m\Phi\psi+
\frac{\tilde{\lambda}}{4m^2}|\psi|^2
\psi=\mu\psi\,.
\end{eqnarray}
Therefore, the system is reduced in the Newtonian limit to the
(non-linear) Schr\"odinger--Poisson system.

In the subsequent sections of this paper, we will concentrate on
the large coupling limit to extract analytic results for compact objects.

%%%%%%%%%%%%%%%%%%%%%%%%%%%%%%%%%%%%%%%%%%%%%%%%%%%%%%%%%%%%%%%%%%%%%%%%%%%
%%%%%%%%%%%%%%%%%%%%%%%%%%%%%%%%%%%%%%%%%%%%%%%%%%%%%%%%%%%%%%%%%%%%%%%%%%%
\section{Large coupling limit and the spherical solution}
\label{sec3}
%%%%%%%%%%%%%%%%%%%%%%%%%%%%%%%%%%%%%%%%%%%%%%%%%%%%%%%%%%%%%%%%%%%%%%%%%%%
%%%%%%%%%%%%%%%%%%%%%%%%%%%%%%%%%%%%%%%%%%%%%%%%%%%%%%%%%%%%%%%%%%%%%%%%%%%
Here, we consider the large coupling limit \cite{colpi}.
We assume the case of $\Lambda\gg 1$, where
\begin{equation}
\Lambda=\frac{\tilde{\lambda}}{8\pi Gm^2}\,.
\end{equation}
In addition, if we introduce the following quantities
\begin{equation}
\mbox{\boldmath $r$}_*=\frac{m}{\sqrt{\Lambda}}\mbox{\boldmath
$r$}\,,\quad
\Psi=\sqrt{\frac{4\pi G\Lambda}{m}}\psi\,,\quad
{\mu_*}=\frac{\mu}{m}\,,
\label{scale}
\end{equation}
the set of equations reduces to the simple form
\begin{eqnarray}
& &\nabla^2_*\Phi=|\Psi|^2\,,\label{Poi}
 \\
& &-\frac{1}{2\Lambda}\nabla^2_*\Psi+\Phi\Psi+
\frac{1}{2}|\Psi|^2
\Psi={\mu_*}\Psi\,,
\label{LCL}
\end{eqnarray}
where $\nabla^2_*$ is the rescaled Laplacian expressed in terms of the
coordinate $\mbox{\boldmath $r$}_*$.

In the limit of $\Lambda\rightarrow\infty$, equation (\ref{LCL})
further reduces to
\begin{equation}
\Phi\Psi+
\frac{1}{2}|\Psi|^2
\Psi={\mu_*}\Psi\,.
\label{LL}
\end{equation}

We can interpret the solutions of (\ref{LL}) as follows.
In the region outside the boson star, the solution is $\Psi\equiv 0$.
In the region inside the boson star, i.e., in the region of $|\Psi|>0$,
the solution is expressed by
\begin{equation}
{\Phi} =\mu_*-\frac{1}{2}\rho\,,
\label{grav}
\end{equation}
where the normalized density function is defined as
\begin{equation}
\rho\equiv|\Psi|^2\,.
\end{equation}
Note that $\mu_*$ indicates the value of the gravitational potential at
the surface of the boson star, where $\rho=0$. If the relation
(\ref{grav}) is substituted in the Poisson equation (\ref{Poi}), we
obtain the linear differential equation
\begin{equation}
\nabla_*^2\rho+2\rho=0\,,
\label{le}
\end{equation}
which is valid in the region inside the boson star, while $\rho=0$
outside the star.%
\footnote{Note that, in the large coupling limit, the boson star has no
asymptotic tail outside the star \cite{Ryan,colpi}.}  %R1 
Because of Eq.~(\ref{ndef}), the solution of
the linear equation (\ref{le}) should be normalized as
\begin{equation}
N=\frac{\sqrt{\Lambda}}{4\pi Gm^2}\int d^3\mbox{\boldmath $r$}_*\,
\rho(\mbox{\boldmath
$r$}_*)\,.
\label{np}
\end{equation}

%R1
Now, we consider the spherically symmetric solution of the system. 
Then, the equation (\ref{le}) can be rewritten as
\begin{equation}
\frac{1}{r_*}\frac{d^2}{dr_*^2}(r_*\rho)+2\rho=0\,,
\end{equation}
where
$r_*=\sqrt{\mbox{\boldmath $r$}_*\cdot\mbox{\boldmath $r$}_*}$.
The normalized solution for $\rho(r_*)$ is
analytically expressed as 
\begin{equation}
\rho(r_*)=\left\{
\begin{array}{cc}\frac{4\pi
Gm^2}{\sqrt{\Lambda}}\frac{N}{\sqrt{2}\pi^2}\frac{\sin\sqrt{2}
r_*}{\sqrt{2}r_*} &\quad (r_*<\pi/\sqrt{2})\\
0 & \quad (r_*>\pi/\sqrt{2})
\end{array}
\right.\,.
\label{sol1}
\end{equation}
The solution for the gravitational potential $\Phi(r_*)$ can be found
by Eq.~(\ref{grav}) for $r_*\le \pi/\sqrt{2}$. We should choose the value
of $\mu_*$ so that $\Phi(r_*)$ matches the Newton potential at
$r_*=\pi/\sqrt{2}$.
Hence, we find
\begin{equation}
\Phi(r_*)=\left\{
\begin{array}{cc}
-\frac{\sqrt{2}}{\pi}\frac{Gm^2N}{\sqrt{\Lambda}}-\frac{2\pi
Gm^2}{\sqrt{\Lambda}}\frac{N}{\sqrt{2}\pi^2}\frac{\sin\sqrt{2}
r_*}{\sqrt{2}r_*} & \quad (r_*<\frac{\pi}{\sqrt{2}})\\
-\frac{Gm^2N}{\sqrt{\Lambda}r_*} &
\quad (r_*>\frac{\pi}{\sqrt{2}})
\end{array}
\right.\,.
\label{sol2}
\end{equation}
Note that $\mu_*$ takes a negative value in general.
%R1

The Newtonian energy $E$ of the system in the
large coupling limit can be expressed from Eq.~(\ref{NE}) as
\begin{equation}
E=\frac{\sqrt{\Lambda}}{4\pi Gm}\int d^3\mbox{\boldmath $r$}_*
\left[\frac{1}{2}({\nabla}_*
\Phi)^2
%+\frac{1}{2\Lambda}|\nabla_*\psi_*|^2
+\Phi |\Psi|^2+
\frac{1}{4}|\Psi|^4
\right]\,.
\label{energy}
\end{equation}
Substituting the solution of Eqs.~(\ref{le}) and (\ref{grav}) into this
equation, we obtain
\begin{equation}
E\cong\frac{\sqrt{\Lambda}}{4\pi Gm}\int d^3\mbox{\boldmath $r$}_*
\left[\frac{1}{2}\Phi \rho+
\frac{1}{4}\rho^2
\right]=\frac{\sqrt{\Lambda}}{4\pi Gm}\int d^3\mbox{\boldmath $r$}_*
\left[\frac{1}{2}\mu_*
\rho\right]=\frac{1}{2}Nm\mu_*=\frac{1}{2}N\mu\,.
\label{sphe}
\end{equation}

The mass of the boson star is given in the present Newtonian scheme by
\begin{equation}
M=Nm+E\,.
\end{equation}
After substituting the solution  (\ref{sol2}) into (\ref{sphe}),
we find that the mass of the spherical boson star becomes
\begin{equation}
M(N)=mN-\frac{Gm^3}{\sqrt{2}\pi\sqrt{\Lambda}}N^2\,.
\end{equation}

Incidentally, if we can vary the value of $N$, the maximum of $M$ occurs
for
\begin{equation}
N=\frac{\pi}{\sqrt{2}}\frac{\sqrt{\Lambda}}{Gm^2}\,.
\end{equation}
The maximum value of $M$ is
\begin{equation}
M_{\rm max}=\frac{\pi}{2\sqrt{2}}\frac{\sqrt{\Lambda}}{Gm}\,,
\end{equation}
%The radius of the boson star is
%\begin{equation}
%R=\frac{\pi}{\sqrt{2}}\frac{\sqrt{\Lambda}}{m}\,.
%\end{equation}
which is supposed to be a typical mass of the boson star in the large
coupling limit. This value for the mass of the boson star is several times
greater than the general relativistic result \cite{colpi}.
This is very similar to the known case for the Newtonian and relativistic
boson stars with no self-interaction \cite{Jetzer}.
We need not consider the boson star with the maximum mass,
especially for explaining the galaxy rotation caused by a single huge
boson star located at the center of the galaxy \cite{TCL,ST,MA}.

In $D$-dimensional spacetime, we take the same forms of the Lagrangian
density
${\cal L}$ and the Hamiltonian density ${\cal H}$ as the
four-dimensional ones. Then, the shape of the equation of motion for
$\rho$ is unchanged; the rescaled Laplacian in the spherical system of
$(D-1)$-dimensional space is replaced by
\begin{equation}
\nabla_*^2=\frac{1}{r_*^{D-2}}\frac{d}{dr_*}r_*^{D-2}\frac{d}{dr_*}\,.
\end{equation}
For a $D$-dimensional boson star, the particle number $N$ is expressed as
\begin{equation}
N=\frac{(\sqrt{\Lambda})^{D-3}}{4\pi Gm^{D-2}}\int d^{D-1}\mbox{\boldmath
$r$}_*\,
\rho(\mbox{\boldmath
$r$}_*)\,.
\end{equation}
By solving the higher-dimensional equation, we find the solution
\begin{equation}
\rho(r_*)=\left\{
\begin{array}{cc}
\frac{4\pi Gm^{D-2}}{A_{D-1}(\sqrt{\Lambda})^{D-3}}
\frac{2^{\frac{D-1}{2}}N}{q^{\frac{D-1}{2}}
J_{\frac{D-1}{2}}(q)}
\frac{J_{\frac{D-3}{2}}(\sqrt{2}
r_*)}{(\sqrt{2}r_*)^{\frac{D-3}{2}}} &\quad (r_*<\frac{q}{\sqrt{2}})\\
0 & \quad (r_*>\frac{q}{\sqrt{2}})
\end{array}
\right.\,,
\label{sold1}
\end{equation}
and
\begin{equation}
\Phi(r_*)=\left\{
\begin{array}{cc}
-\frac{4\pi
Gm^{D-2}N}{A_{D-1}(\sqrt{\Lambda})^{D-3}}
\frac{2^{\frac{D-3}{2}}}{(D-3)q^{D-3}}-\frac{2\pi
Gm^{D-2}}{A_{D-1}(\sqrt{\Lambda})^{D-3}}
\frac{2^{\frac{D-1}{2}}N}{q^{\frac{D-1}{2}}
J_{\frac{D-1}{2}}(q)}
\frac{J_{\frac{D-3}{2}}(\sqrt{2}
r_*)}{(\sqrt{2}r_*)^{\frac{D-3}{2}}} & \quad (r_*<\frac{q}{\sqrt{2}})\\
-\frac{4\pi
Gm^{D-2}N}{A_{D-1}(\sqrt{\Lambda})^{D-3}}
\frac{1}{(D-3)r_*^{D-3}}
&
\quad (r_*>\frac{q}{\sqrt{2}})
\end{array}
\right.\,,
\label{sold2}
\end{equation}
where
$A_{D-1}\equiv\frac{2\pi^{\frac{D-1}{2}}}{\Gamma\left(\frac{D-1}{2}\right)}$
, $J_n(z)$ is the Bessel function of the
first kind, and $q$ is the first non-trivial zero of
$J_{\frac{D-3}{2}}(x)$, i.e.,
$J_{\frac{D-3}{2}}(q)=0$.
%R1
As in the case of $D=4$, we have chosen the value of $\mu_*$
so that $\Phi(r_*)$ matches the gravitational potential in vacuum
outside the star. 

The Newtonian energy is the same as that in four dimensions:
\begin{equation}
E=\frac{1}{2}Nm\mu_*=\frac{1}{2}N\mu\,.
\end{equation}
Thus, the mass of the spherical boson star is given by
\begin{equation}
M(N)=Nm-\frac{4\pi
Gm^{D-1}}{2A_{D-1}(\sqrt{\Lambda})^{D-3}}\frac{2^{\frac{D-3}{2}}}{(D-3)q^{D-3}}N^2\,.
\end{equation}
The maximum of $M$ occurs when
\begin{equation}
N=\frac{A_{D-1}(\sqrt{\Lambda})^{D-3}}{4\pi
Gm^{D-2}}
\frac{(D-3)q^{D-3}}{2^{\frac{D-3}{2}}}\,,
\end{equation}
and the maximum mass is
\begin{equation}
M_{\rm max}=\frac{A_{D-1}(\sqrt{\Lambda})^{D-3}}{4\pi
Gm^{D-3}}
\frac{(D-3)q^{D-3}}{2^{\frac{D-1}{2}}}\,.
\end{equation}

%%%%%%%%%%%%%%%%%%%%%%%%%%%%%%%%%%%%%%%%%%%%%%%%%%%%%%%%%%%%%%%%%%%%%%%%%%%
%%%%%%%%%%%%%%%%%%%%%%%%%%%%%%%%%%%%%%%%%%%%%%%%%%%%%%%%%%%%%%%%%%%%%%%%%%%
\section{Rotating Newtonian boson stars with large
self-interaction}
\label{sec4}
%%%%%%%%%%%%%%%%%%%%%%%%%%%%%%%%%%%%%%%%%%%%%%%%%%%%%%%%%%%%%%%%%%%%%%%%%%%
%%%%%%%%%%%%%%%%%%%%%%%%%%%%%%%%%%%%%%%%%%%%%%%%%%%%%%%%%%%%%%%%%%%%%%%%%%%

Now we turn to the case of $D=4$ again and consider rotating boson stars.
We assume their axial symmetry and equatorial symmetry.

For a stationary rotating boson star, we set
\cite{R1S,R1SM1,R1SM2,Ryan,HRR}
\begin{equation}
\Psi\rightarrow\Psi(r_*, \theta)\,e^{is\varphi}\,,
\end{equation}
where $\varphi$ indicates the polar angle and an integer $s$ corresponds
to the angular momentum. Substituting this ansatz, Eq.~(\ref{LCL}) reads
\begin{equation}
-\frac{1}{2\Lambda}\left(\frac{1}{r_*}\frac{\partial^2}{\partial
r_*^2}r_*\Psi+\frac{1}{r_*^2\sin\theta}\frac{\partial}{\partial\theta}
\sin\theta\frac{\partial}{\partial\theta}\Psi-\frac{s^2}{r_*^2\sin^2\theta}\Psi\right)+\Phi\Psi+
\frac{1}{2}|\Psi|^2
\Psi={\mu_*}\Psi\,.
\end{equation}

We assume the large coupling limit $\Lambda\rightarrow\infty$ as well as
the rapid rotation of the phase of the scalar field such that $s_*^2\equiv
s^2/\Lambda$ takes a finite value. Then, the equation becomes
\begin{equation}
\left[\Phi+
\frac{1}{2}\left(|\Psi|^2
+\frac{s_*^2}{r_*^2\sin^2\theta}\right)-{\mu_*}\right]\Psi=0\,.
\end{equation}
If the relation for $\Psi\ne 0$ is substituted into the Poisson equation,
we obtain the following inhomogeneous differential equation:
\begin{equation}
\nabla_*^2f+2f=
\frac{2s_*^2}{r_*^2\sin^2\theta}\,,
\label{le2}
\end{equation}
where
\begin{equation}
f(r_*,\theta)\equiv \rho
+\frac{s_*^2}{r_*^2\sin^2\theta}\,,
\end{equation}
with $\rho\equiv|\Psi|^2$.

The particular solution of (\ref{le2}) is given by \cite{wolfram}
\begin{equation}
s_*^2f_p(x)\equiv -\frac{s_*^2\pi}{2}Y_0(\sqrt{2}x)
G_{13}^{20}\left(\frac{x^2}{2}\left|
\begin{array}{c}
1 \\
0, 0, 0
\end{array}\right.
\right)\,,
\label{fpx}
\end{equation}
where $x\equiv r_*\sin\theta$, $Y_n(z)$ is the Bessel function of the
second kind and
$G_{pq}^{mn}$ is the Meijer $G$ function.

The general solution to Eq.~(\ref{le2}) with axial and equatorial
symmetries ($\rho(r_*,\theta)=\rho(r_*,\pi-\theta)$) is expressed by the
particular solution
$s_*^2f_p(x)$ plus the linear combination
\begin{equation}
\sum_{n=0}^\infty j_{2n}(\sqrt{2}r_*)P_{2n}(\cos\theta)
\,,
\end{equation}
where $j_l(z)$ is the spherical Bessel function%
\footnote{Note that
$j_0(z)=\sqrt{\frac{\pi}{2z}}J_{1/2}(z)=\frac{\sin z}{z}$.}
and $P_n(x)$ is the Legendre polynomial of the $n$-th order, or
\begin{equation}
\int dk \left[\alpha(k)J_0(\sqrt{2-k^2}x)+
\beta(k)Y_0(\sqrt{2-k^2}x)\right]\cos kz\,,
\end{equation}
where $z\equiv r_*\cos\theta$.

To simplify our analysis, we wish to use the minimal number of
parameters throughout the present paper.
The radius of the boson star is thought to be an important physical
parameter.
Hence, we first consider a simple
ansatz that $\rho$ is given by
\begin{equation}
A\frac{\sin\sqrt{2}(r_*-h)}{\sqrt{2}r_*}+s_*^2\left(
f_p(x)-\frac{1}{x^2}\right)-s_*^2\left(
f_p(R_*)-\frac{1}{R_*^2}\right)\frac{Y_0(\sqrt{2}x)}{Y_0(\sqrt{2}R_*)}\,,
\end{equation}
where the scale factor $A$ is a dimensionless constant if its numerical
value is positive; otherwise,
$\rho=0$. Here, the equatorial radius of the boson star is given by
$R_*=\frac{\pi}{\sqrt{2}}+h$.

\begin{figure}[ht]
\centering
\includegraphics[height=3.5cm]
{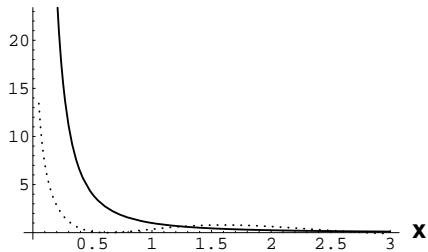}
\caption{%
$f_p(x)$ (the dotted line) and $1/x^2$ (the solid line).
}
\label{fig1}
\end{figure}

As a bolder approximation, we will omit $f_p$ in the previous ansatz.
The special solution $f_p(x)$ behaves logarithmic in the vicinity of the
origin. The term proportional to $1/x^2$ makes $\rho$ vanish more rapidly
near the origin (Fig.~\ref{fig1}).
%%%%%%%%%%%%%%%%%%%%%%%%%%%
%\begin{wrapfigure}{r}{5cm}
%\end{wrapfigure}
%%%%%%%%%%%%%%%%%%%%%%%%%%%
We can therefore omit $f_p(x)$ to make numerical integration efficient,
and we finally adopt the following approximation for $\rho$:
\begin{equation}
A\frac{\sin\sqrt{2}(r_*-h)}{\sqrt{2}r_*}-s_*^2\left(
\frac{1}{x^2}
-\frac{1}{R_*^2}\frac{Y_0(\sqrt{2}x)}{Y_0(\sqrt{2}R_*)}\right)\,,
\label{assumption}
\end{equation}
if its value is positive.

The total particle number is still given by (\ref{np}), as in the
non-rotating case, while the Newtonian energy of the rotating boson star
becomes 
\begin{equation}
E=\frac{\sqrt{\Lambda}}{4\pi Gm}\int d^3\mbox{\boldmath $r$}_*
\left[\frac{1}{2}\rho\left(\mu_*+\frac{s_*^2}{2x^2}\right) \right]\,,
\end{equation}
which is deduced from (\ref{NE}) and field equations.

Now, we solve the Poisson equation (\ref{le}) with the $\rho$ found above
in order to evaluate the value of $\mu_*$.
The solution for the Poisson equation can be expressed as follows:
\begin{equation}
\Phi(\mbox{\boldmath $r$}_*)=\int G(\mbox{\boldmath
$r$}_*-\mbox{\boldmath
$r$}_*{}')\rho(\mbox{\boldmath
$r$}_*{}') d^3\mbox{\boldmath
$r$}_*{}'\,,
\end{equation}
where the Green function $G$ satisfies
\begin{equation}
\nabla_*^2G(\mbox{\boldmath $r$}_*-\mbox{\boldmath
$r$}_*{}')=\delta^3(\mbox{\boldmath $r$}_*-\mbox{\boldmath
$r$}_*{}')\,.
\end{equation}
The Green function satisfying the condition $G\rightarrow 0$ at the
spatial infinity is well known and is given by
\begin{equation}
G(\mbox{\boldmath $r$}_*-\mbox{\boldmath
$r$}_*{}')=-\frac{1}{4\pi|\mbox{\boldmath $r$}_*-\mbox{\boldmath
$r$}_*{}'|}\,,
\end{equation}
or
\begin{equation}
-\frac{1}{4\pi|\mbox{\boldmath $r$}_*-\mbox{\boldmath
$r$}_*{}'|}=-\sum_{l=0}^\infty
\sum_{m=-l}^l
\frac{1}{2l+1}\frac{r^l_<}{r^{l+1}_>}Y^*_{lm}(\theta',\varphi')
Y_{lm}(\theta,\varphi)\,,
\end{equation}
where $r_<=\min(r_*,r_*{}')$, $r_>=\max(r_*,r_*{}')$, and
$Y_{lm}(\theta, \varphi)$ is the spherical harmonic function.

Because the density distribution has been assumed to be axially and
equatorially symmetric, the gravitational potential $\Phi$ outside the
boson star can be obtained as
\begin{equation}
\Phi(x, z)=-\int_0^\infty\int_0^\infty\rho(x', z')
\frac{2K\left(-\frac{4xx'}{(x-x')^2+(z-z')^2}\right)}{\pi\sqrt{(x-x')^2+(z-z')^2}}
\,x' dx' dz'\,,
\label{p1}
\end{equation}
where $K(m)$ is the complete elliptic integral of the first kind.
%, or
%\begin{equation}
%\Phi(r_*, \theta)=-\frac{1}{2}\sum_{n=0}^\infty
%\frac{P_{2n}(\theta)}{r_*^{2n+1}}\int_0^\infty\int_0^\pi{r_*'}^{2n}P_{2n}(\theta')
%\rho(r_*',\theta'){r_*'}^2dr_*'\sin\theta' d\theta'\,.
%\label{p2}
%\end{equation}
%%%%%%%%%%%%%%%%%%%%%%%%%%%
%In Eq.~(\ref{p2}), we have used an integration formula
%\begin{equation}
%\int_0^{2\pi}\frac{d\varphi}{\sqrt{x^2+{x'}^2-2xx'\cos\varphi+(z-z')^2}}=
%\frac{4K\left(-\frac{4xx'}{(x-x')^2+(z-z')^2}\right)}{\sqrt{(x-x')^2+(z-z')^2}}
%\,.
%\end{equation}
%%%%%%%%%%%%%%%%%%%%%%%%%%%
By numerically integrating (\ref{p1}), we can obtain the value
of
$\mu_*$ as
\begin{equation}
\mu_*=\Phi(R_*,0)\,.
\end{equation}

Now, we can numerically calculate the Newtonian energy $E$.
The shape of the boson star is determined by $h$
 and
$s_*^2/A$ in the form (\ref{assumption}).%
\footnote{Note that the unit of $h$ is $\sqrt{\Lambda}/m$ (see
Eq.~(\ref{scale})).}
We can evaluate the normalized Newtonian binding energy 
\begin{equation}
\frac{\sqrt{\Lambda}}{4\pi Gm^3}\frac{E}{N^2}\,,
\end{equation}
which is apparently independent of the scale factor $A$.

%%%%%%%%%%%%%%%%%%%%%%%%%%%
\begin{table}[htb]
\begin{center}
{\footnotesize
\begin{tabular}{|c|c|ccccccc|}\hline
&&&&\multicolumn{3}{c}{$s_*^2/A$}&&\\ \hline
& & 0.1 & 0.3 & 0.5 & 0.7 & 0.9 & 1.1 & 1.3 \\ \hline
&0.00 & -0.0158283 & -0.0116258 & -0.00630139 & -0.000286516 & 
   0.00780719 & 0.0163551 & 0.0256774 \cr 
&0.05 &-0.0157169 & -0.0121735 & 
    -0.00791623 & -0.00358682 & 0.00244601 & 0.00825089 & 
   0.0142766 \cr 
&0.10 & -0.0155833 & -0.0126021 & -0.0092137 & -0.00551656 & 
    -0.00165121 & 0.00226248 & 0.00613015 \cr 
 &0.15 &-0.0154328 & 
    -0.0129336 & -0.0102595 & -0.00751525 & -0.00480722 & 
    -0.00221123 & 0.000229342 \cr 
$h$&0.20 & -0.0152694 & -0.0131877 & 
    -0.0111047 & -0.00910385 & -0.00724985 & -0.00557194 & 
    -0.00407807 \cr
&0.25 &-0.015097 & -0.0133802 & -0.0117865 & -0.0103671 & 
    -0.00913752 & -0.00809426 & -0.0072192 \cr
&0.30 &-0.0149176 & 
    -0.0135201 & -0.0123334 & -0.0113593 & -0.0105769 & -0.00995592 & 
    -0.00946259 \cr
&0.35 &-0.0147352 & -0.0136173 & -0.0127593 & 
    -0.0121104 & -0.0116216 & -0.0112503 & -0.0109625 \cr
&0.40 &-0.0145493 & 
    -0.0136709 & -0.0130539 & -0.0126215 & -0.0123162 & -0.0120917 & 
    -0.0119222 \\ \hline
\end{tabular}
}
\caption{The values of $(\sqrt{\Lambda}/(4\pi Gm^3))E/N^2$ for the
four-dimensional boson star.%G4
}
\label{taen2}
\end{center}
\end{table}
%%%%%%%%%%%%%%%%%%%%%%%%%%%

Table~\ref{taen2} lists the values of $(\sqrt{\Lambda}/(4\pi Gm^3))E/N^2$
for the four-dimensional boson star as a function of $h$ and
$s_*^2/A$. If $(\sqrt{\Lambda}/(4\pi Gm^3))E/N^2$ is negative, the
solution is considered to be energetically stable.

%%%%%%%%%%%%%%%%%%%%%%%%%%%
%\begin{wrapfigure}{r}{5cm}
\begin{figure}[ht]
\centering
\includegraphics[height=5cm]
{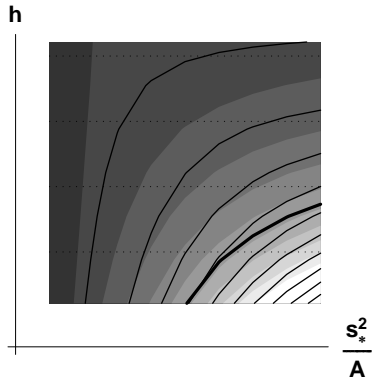}
\caption{%
The normalized Newtonian energy $E/N^2$ for the boson star in four
dimensions shown by gray shades. On the thick line, the
energy vanishes ($E=0$). The dotted lines indicate contours of equal radii
while the solid lines indicate contours of equal
$s_*^2/N$. For the parameter region, please see the text.}
\label{figen2}
\end{figure}
%\end{wrapfigure}
%%%%%%%%%%%%%%%%%%%%%%%%%%%

Figure~\ref{figen2} shows this normalized Newtonian binding energy
$(\sqrt{\Lambda}/(4\pi Gm^3))E/N^2$ in a contour plot. The horizontal axis
indicates
$s_*^2/A=\{0.1,1.3\}$, while the vertical axis indicates
$h=\{0.0,0.4\}$. A darker region corresponds to a lower energy.
Each solid line shows the line on which the value
$\frac{\sqrt{\Lambda}}{4\pi G m^2}\frac{s_*^2}{N}=\frac{s^2}{4\pi
G m^2 \sqrt{\Lambda}N}$ is constant.

For a small angular momentum or a small $s^2/N$, a lower $h$ yields
a lower $E/N^2$. Thus, the equatorial radius of the boson star is
$R_*=\pi/\sqrt{2}$ for a small
$s^2/N$. For a large $s^2/N$, the binding energy $E$ becomes lower at a
finite $h$. 
Under the condition that the conserved quantities, the particle number and
angular momentum of the boson star, are fixed, the configuration with
lower energy is considered to be realized. Therefore, for a fixed
$N$, the equatorial radius of the rapidly rotating boson star
increases with a higher angular momentum. This qualitative behavior
corresponds to the result in the relativistic system studied by Ryan
\cite{Ryan} in the large-coupling limit.

%%%%%%%%%%%%%%%%%%%%%%%%%%%
%\begin{wrapfigure}{r}{5cm}
\begin{figure}[ht]
\centering
\includegraphics[height=4cm]
{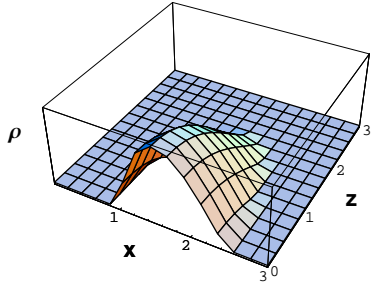}
\caption{%
A profile of the Newtonian rotating boson star
with the parameters $h=0.25$ and $s_*^2/A=0.8$.
}
\label{figsh}
\end{figure}
%\end{wrapfigure}
%%%%%%%%%%%%%%%%%%%%%%%%%%%

We show a typical profile of the rotating boson star in Fig.~\ref{figsh},
where the scale of the vertical axis is taken arbitrarily since the shape
is independent of the scale factor $A$. %R1

%%%%%%%%%%%%%%%%%%%%%%%%%%%
\begin{table}[htb]
\begin{center}
\begin{tabular}{|c|c|ccccccc|}\hline
&&&&\multicolumn{3}{c}{$s_*^2/A$}&&\\ \hline
& & 0.1 & 0.3 & 0.5 & 0.7 & 0.9 & 1.1 & 1.3 \\ \hline
&0.00 &61.5431 & 26.9759 & 15.9985 & 10.7511 & 7.79774 & 5.96767 & 
   4.75667 \cr
&0.05 &67.3542 & 30.3416 & 18.5505 & 12.867 & 9.63173 & 
   7.60095 & 6.23663 \cr
&0.10 &73.5307 & 34.0246 & 21.426 & 15.3135 & 
   11.8044 & 9.5755 & 8.05857 \cr
 &0.15&80.0819 & 38.0562 & 24.6637 & 
   18.1383 & 14.3593 & 11.9336 & 10.2657 \cr 
$h$&0.20 &87.0118 & 42.4584 & 
   28.2936 & 21.3601 & 17.3033 & 14.667 & 12.8201 \cr
&0.25 &94.3121 & 
   47.208 & 32.4701 & 24.9027 & 20.5056 & 17.5532 & 15.3778 \cr
&0.30 &101.936 & 52.2142 & 36.3944 & 28.3421 & 23.2303 & 19.4039 & 
   16.4532 \cr
&0.35 &109.702 & 56.997 & 39.5987 & 29.6371 & 22.7876 & 
   18.0043 & 14.5134 \cr
&0.40 &117.032 & 59.4375 & 37.0806 & 24.1276 & 
   16.0474 & 10.5961 & 6.67022 \\ \hline
\end{tabular}
\caption{The value of $4\pi G\sqrt{\Lambda}m |Q|/s^2$ for the
four-dimensional boson star.%G4
}
\label{taq}
\end{center}
\end{table}
%%%%%%%%%%%%%%%%%%%%%%%%%%%

Next, we list the values of $4\pi G\sqrt{\Lambda}m |Q|/s^2$, the ratio
of the quadrupole moment to $s^2$ of  the boson star, in Table~\ref{taq}. 
The quadrupole moment $Q$ is given by
\begin{equation}
Q=\frac{\sqrt{\Lambda}}{4\pi Gm}\int d^3\mbox{\boldmath $r$}_*\,
\rho(\mbox{\boldmath
$r$}_*)\left(3z^2-|\mbox{\boldmath $r$}_*|^2\right)\,.
\label{Q}
\end{equation}
%%%%%%%%%%%%%%%%%%%%%%%%%%%
%\begin{wrapfigure}{r}{5cm}
\begin{figure}[ht]
\centering
\includegraphics[height=5cm]
{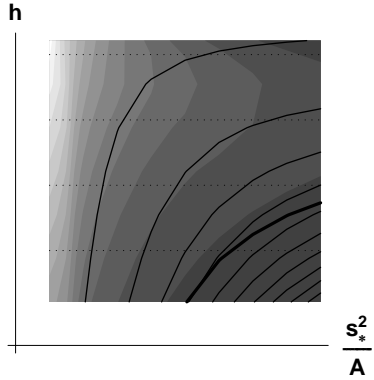}
\caption{%
The ratio of the quadrupole moment to $s^2$, $4\pi G\sqrt{\Lambda}m
|Q|/s^2$ (see (\ref{Q})), of the boson star shown by gray
shades.}
\label{figqs2}
\end{figure}
%\end{wrapfigure}
%%%%%%%%%%%%%%%%%%%%%%%%%%%
The ratio of the quadrupole moment to $s^2$ of the boson star is shown in
Fig.~\ref{figqs2} as a contour plot. 

The parameter region is the same as in Fig.~\ref{figen2}.
From this plot, we find the following.
For a small $s^2/N$, the ratio is small; 
when the radius of the boson star increases, the ratio increases.
This qualitative behavior coincides with the relativistic result obtained
by Ryan \cite{Ryan} in the large-coupling limit.
Because the quadrupole moment is sensitive to the detail of the shape,
it is difficult to study on the qualitative correspondence to the previous
result in the present analysis with the minimal number of
parameters.

%%%%%%%%%%%%%%%%%%%%%%%%%%%%%%%%%%%%%%%%%%%%%%%%%%%%%%%%%%%%%%%%%%%%%%%%%%%
%%%%%%%%%%%%%%%%%%%%%%%%%%%%%%%%%%%%%%%%%%%%%%%%%%%%%%%%%%%%%%%%%%%%%%%%%%%
\section{Interlude: on boson stars in the Kaluza-Klein background}
\label{sec5}
%%%%%%%%%%%%%%%%%%%%%%%%%%%%%%%%%%%%%%%%%%%%%%%%%%%%%%%%%%%%%%%%%%%%%%%%%%%
%%%%%%%%%%%%%%%%%%%%%%%%%%%%%%%%%%%%%%%%%%%%%%%%%%%%%%%%%%%%%%%%%%%%%%%%%%%

In this section, we consider a boson-star solution in the Kaluza-Klein
background, i.e., a boson star in spacetime with a compact extra
dimension.

We denote the coordinate of the extra space as $y$. This dimension is
assumed to be a circle, the circumference of which is set as $L$.
Thus, the coordinate can be considered periodic by identifying $y+L\sim
y$.

As in previous sections, we rescale the length scale as
\begin{equation}
y_*=\frac{m}{\sqrt{\Lambda}}y\,,\quad L_*=\frac{m}{\sqrt{\Lambda}}L\,.
\end{equation}

The Lagrangian density is assumed to be of the same form as in the
previous case.  Accordingly, a non-zero $\rho$ satisfies the differential
equation
\begin{equation}
\left(\frac{1}{r_*^2}\frac{\partial}{\partial
r_*}{r_*^2}\frac{\partial}{\partial r_*}+\frac{\partial^2}{\partial y_*^2}
+2\right)\rho(r_*,y_*)=0\,,
\end{equation}
if the spherical symmetry in three-dimensional space is assured.
For a sufficiently small $L_*$, a solution for a boson star can be
described by
\begin{equation}
\rho=\max\left[A\frac{\sin\sqrt{2}r_*}{\sqrt{2}r_*},\,0\right]\,,
\end{equation}
where the scale factor $A$ is a positive constant. This solution is the
same as the four-dimensional solution; this configuration may appear like
a `boson string' in four-dimensional space.

On the other hand, for a finite length $L$, we conjecture the
following solution for
$\rho$:
\begin{equation}
\rho=\max\left[A\left(\frac{\sin\sqrt{2}r_*}{\sqrt{2}r_*}+\delta\frac{\sin\sqrt{2-\frac{4\pi^2}{L_*^2}}r_*}{\sqrt{2}r_*}
\cos\frac{2\pi}{L_*}y_*\right),\,0\right]\,,
\end{equation}
if $\sqrt{2}\pi<L_*<2\sqrt{2}\pi$. Here, $\delta$ is a constant.
Note that at the special location $y_*=L_*/4$, the surface of the boson
star is located at
$r_*=R_*=\pi/\sqrt{2}$.

Because of the periodicity in the extra coordinate $y_*$, the Green
function for the Poisson equation defined in the Kaluza-Klein background
is given by
\begin{eqnarray}
G(\Delta\mbox{\boldmath
$r$}_*, \Delta
y_*)&=&-\frac{1}{4\pi^2}\sum_{n=-\infty}^\infty
\frac{1}{|\Delta\mbox{\boldmath
$r$}_*|^2+(\Delta y_*+ nL_*)^2}\nonumber \\
&=& -\frac{\sinh (2\pi |\Delta\mbox{\boldmath
$r$}_*|/L_*)}{2L_*^2(2\pi|\Delta\mbox{\boldmath
$r$}_*|/L_*)[\cosh (2\pi |\Delta\mbox{\boldmath
$r$}_*|/L_*)-\cos (2\pi y_*/L_*)]}\,,
\end{eqnarray}
where $|\Delta\mbox{\boldmath
$r$}_*|=|\mbox{\boldmath
$r$}_*-\mbox{\boldmath
$r$}_*'|$ and $\Delta y_*=y_*-y_*'$.

Using the assumption that $\rho$ depends on $r_*$ and $y_*$, and has the
symmetry $y_*\leftrightarrow -y_*$, we obtain the gravitational
potential, %
%as in the form of an infinite sum:
%\begin{eqnarray}
%& &\Phi(r_*, y_*)=\int_0^{L_*}dy_*'\int d^3\mbox{\boldmath
%$r$}_*'  G(\Delta\mbox{\boldmath
%$r$}_*, \Delta
%y_*)\rho(r_*, y_*)\nonumber \\
%& &=-\frac{1}{L_*r_*}\int_0^{L_*}dy_*'\int_0^\infty {r_*'}^2dr_*'
%\rho(r_*',y_*')\nonumber \\
%& &-2\sum_{l=1}^\infty\frac{e^{-(2\pi l r_*/L_*)}}{L_*r_*}\cos\frac{2\pi
%l y_*}{L_*}
%\int_0^{L_*}dy_*'\int_0^\infty {r_*'}^2dr_*'\cos\frac{2\pi l y_*'}{L_*}
%\cdot\frac{\sinh(2\pi l r_*'/L_*)}{2\pi l r_*'/L_*}\rho(r_*',y_*')\,.
%\end{eqnarray}
with an approximation of taking only the longest-wave mode, which is
equivalent to neglecting the
$O((e^{-(2\pi  r_*/L_*)})^2)$ terms, as
\begin{eqnarray}
& &\Phi(r_*, y_*)=\int_0^{L_*}dy_*'\int d^3\mbox{\boldmath
$r$}_*'  G(\Delta\mbox{\boldmath
$r$}_*, \Delta
y_*)\rho(r_*, y_*)\nonumber \\
& &\approx -\frac{1}{L_*r_*}\int_0^{L_*}dy_*'\int_0^\infty
{r_*'}^2dr_*'
\rho(r_*',y_*')\nonumber \\
& &-2\frac{e^{-(2\pi  r_*/L_*)}}{L_*r_*}\cos\frac{2\pi
 y_*}{L_*}
\int_0^{L_*}dy_*'\int_0^\infty {r_*'}^2dr_*'\cos\frac{2\pi  y_*'}{L_*}
\cdot\frac{\sinh(2\pi  r_*'/L_*)}{2\pi  r_*'/L_*}\rho(r_*',y_*')\,.
\end{eqnarray}
Thus, at the lowest order, we find
\begin{equation}
\Phi(r_*,L_*/4)\approx-\frac{
Gm^{3}}{\Lambda L_*r_*}N\,,
\end{equation}
where 
\begin{equation}
N=\frac{\Lambda}{4\pi Gm^{3}}\int_0^{L_*}dy_*\int d^3\mbox{\boldmath
$r$}_*\rho(\mbox{\boldmath
$r$}_*, y_*)\,.
\end{equation}
Thus, the Newtonian energy of the boson star becomes
\begin{equation}
E=\frac{1}{2}\mu_*N=\frac{1}{2}\Phi(R_*,L_*/4)N\approx-\frac{
Gm^{3}}{\sqrt{2}\pi
\Lambda L_*}N^2\,,
\end{equation}
up to $O(e^{-4\pi  R_*/L_*})$.

We conclude that the change in energy caused by the variation of the
highest wavelength in the direction of the extra dimension is very small
and is independent of the magnitude of its amplitude $\delta$, when only
the longest-wave mode in the gravitational potential is taken into
consideration.

Although further independent analysis is needed for clarifying the
instability in this boson string, the instability in the
Kaluza-Klein background qualitatively found here reminds us of the
Gregory--Laflamme instability in black strings \cite{GL}. 

%%%%%%%%%%%%%%%%%%%%%%%%%%%%%%%%%%%%%%%%%%%%%%%%%%%%%%%%%%%%%%%%%%%%%%%%%%%
%%%%%%%%%%%%%%%%%%%%%%%%%%%%%%%%%%%%%%%%%%%%%%%%%%%%%%%%%%%%%%%%%%%%%%%%%%%
\section{Newtonian boson stars with large self-coupling in
five dimensions}
\label{sec6}
%%%%%%%%%%%%%%%%%%%%%%%%%%%%%%%%%%%%%%%%%%%%%%%%%%%%%%%%%%%%%%%%%%%%%%%%%%%
%%%%%%%%%%%%%%%%%%%%%%%%%%%%%%%%%%%%%%%%%%%%%%%%%%%%%%%%%%%%%%%%%%%%%%%%%%%

In this section, we consider a rotating boson star in five dimensions.
We consider a system governed by the same form of the Lagrangian
density as in the previous sections.
The coordinates in the five-dimensional spacetime are assumed to be
\begin{equation}
ds^2=-dt^2+dr^2+r^2(d\theta^2+\sin^2\theta
d\varphi_1^2+\cos^2\theta d\varphi_2^2)\,,
\end{equation}
where $0\le\theta\le\pi/2$, $0\le\varphi_1<2\pi$ and $0\le\varphi_2<2\pi$.
After rescaling and redefining the coordinates, the line element
becomes
\begin{eqnarray}
ds^2&=&-dt^2+\frac{\Lambda}{m^2}\left\{dr_*^2+r_*^2(d\theta^2+\sin^2\theta
d\varphi_1^2+\cos^2\theta d\varphi_2^2)\right\}\nonumber \\
&=&-dt^2+\frac{\Lambda}{m^2}\left(dx^2+dy^2+x^2d\varphi_1^2+y^2d\varphi_2^2\right)\,,
\end{eqnarray}
where $x=r_*\sin\theta$ and $y=r_*\cos\theta$.

Here, we assume
\begin{equation}
\Psi\rightarrow\Psi(r_*, \theta)\,e^{is_1\varphi_1+is_2\varphi_2}\,,
\end{equation}
and define $s_{1*}\equiv s_1/\sqrt{\Lambda}$ and $s_{2*}\equiv
s_2/\sqrt{\Lambda}$.
As in Sec.~\ref{sec4}, the large coupling limit yields the
following relation in this case:
\begin{equation}
\Phi+
\frac{1}{2}\left(\rho
+\frac{s_{1*}^2}{x^2}
+\frac{s_{2*}^2}{y^2}\right)-{\mu_*}=0\,,
\end{equation}
for the region of $\rho=|\Psi|^2>0$.
The differential equation in this case reads
\begin{equation}
\nabla_*^2f+2f=
\frac{2s_{1*}^2}{r_*^2\sin^2\theta}
+\frac{2s_{2*}^2}{r_*^2\cos^2\theta}\,,
\label{le5}
\end{equation}
where
\begin{equation}
f(r_*,\theta)\equiv \rho
+\frac{s_{1*}^2}{r_*^2\sin^2\theta}
+\frac{s_{2*}^2}{r_*^2\cos^2\theta}\,,
\end{equation}
in the region of $\rho>0$.
For simplicity, we will consider only the simplest case with
$s_1=s_2$  in this paper.

We conjecture that the boson star is spherical in the limit of no
rotation. Since the Laplacian for the spherical body reads
\begin{equation}
\nabla_*^2=\frac{1}{r_*^3}\frac{\partial}{\partial
r_*}r_*^3\frac{\partial}{\partial r_*}
\,,
\end{equation}
the spherical solution of Eq.~(\ref{le5}) with $s_{1*}=s_{2*}=0$ can be
written by
\begin{equation}
\frac{A}{\sqrt{2}r_*}\left\{J_1(\sqrt{2}r_*)-
\frac{J_1\left(\sqrt{2}\left(\frac{q}{\sqrt{2}}+h\right)\right)}%
{Y_1\left(\sqrt{2}\left(\frac{q}{\sqrt{2}}+h\right)\right)}
Y_1(\sqrt{2}r_*)\right\}
\,,
\end{equation}
where $A$ and $h$ are constants. $q\approx 3.83171$ is the first zero of
$J_1(x)$.

Because the Laplacian can also be expressed as 
\begin{equation}
\nabla^2_*=\frac{1}{x}\frac{\partial}{\partial x}x\frac{\partial}{\partial
x} +\frac{1}{y}\frac{\partial}{\partial y}y\frac{\partial}{\partial y}
+\frac{1}{x^2}\frac{\partial^2}{\partial\varphi_1^2}
+\frac{1}{y^2}\frac{\partial^2}{\partial\varphi_2^2}\,,
\end{equation}
the special solution of (\ref{le5}) is determined to be
$s_{1*}^2f_p(x)+s_{2*}^2f_p(y)$, where $f_p(x)$ is defined by (\ref{fpx}).
We neglect the contribution of $f_p$ to the
solution, as in Sec.~\ref{sec4}.
In Sec.~\ref{sec4}, we parametrize the equatorial radius, but in the
present case, because there are `holes' in the direction of the $x$ and
$y$ axes, we abandon the tuning of the radius as an input parameter.

For simplicity, we consider the case with
$s_{1*}=s_{2*}\equiv s_*$. Hence, the boson star has spherical
symmetry in the limit of no rotation, and we take an approximated
solution:
\begin{eqnarray}
& &\frac{A}{\sqrt{2}r_*}\left\{J_1(\sqrt{2}r_*)-
\frac{J_1\left(\sqrt{2}R_*\right)}%
{Y_1\left(\sqrt{2}R_*\right)}
Y_1(\sqrt{2}r_*)\right\}-s_{*}^2\left(
\frac{1}{x^2}
+\frac{1}{y^2}\right)\,,
\label{hd}
\end{eqnarray}
where $R_*\equiv q/\sqrt{2}+h$, provided that the value of (\ref{hd})
takes a positive value. Note that $R_*$ is not the radius of the boson
star.

The physical quantities of the boson star can be derived as in the
four-dimensional case. The particle number of the boson star is expressed
by
\begin{equation}
N=\frac{\Lambda}{4\pi Gm^{3}}\int d^{4}\mbox{\boldmath
$r$}_*\,
\rho(\mbox{\boldmath
$r$}_*)\,,
\end{equation}
while the Newtonian binding energy is given by
\begin{equation}
E=\frac{\Lambda}{4\pi Gm^{2}}\int d^{4}\mbox{\boldmath
$r$}_*\,
\left[\frac{1}{2}\rho\left(\mu_*+\frac{s_{1*}^2}{2x^2}
+\frac{s_{2*}^2}{2y^2}\right)\right]\,,
\end{equation}
for arbitrary values of $s_{1*}$ and $s_{2*}$.

The value of the gravitational potential at the boson star surface,
$\mu_*$, should be obtained using the Green function in the flat
four-dimensional space. The Green function, which
asymptotically vanishes, is expressed as
\begin{eqnarray}
G(\mbox{\boldmath
$r$}_*,\mbox{\boldmath
$r$}_*')&=&-\frac{1}{4\pi^2}\frac{1}{|\mbox{\boldmath
$r$}_*-\mbox{\boldmath
$r$}_*'|^2}\nonumber \\
&=&-\frac{1}{4\pi^2}\frac{1}{
x^2+{x'}^2-2xx'\cos(\varphi_1-\varphi_1')+
y^2+{y'}^2-2yy'\cos(\varphi_2-\varphi_2')}\,.
\end{eqnarray}
Owing to the two axial symmetries of the boson star configuration,
we wish to integrate the Green function over two polar coordinates.
There is, however, no known compact expression for the integration,
contrary to the case of the three spatial dimensions in Sec.~\ref{sec4}.
Thus, we take a further approximation.

For $r_*'\ll r_*$, the integration over the polar coordinates yields
\begin{eqnarray}
& &\frac{1}{(2\pi)^2}\int_0^{2\pi}d\varphi_1'\int_0^{2\pi}d\varphi_2'
\,G(\mbox{\boldmath
$r$}_*,\mbox{\boldmath
$r$}_*')\nonumber \\
& &=-\frac{1}{4\pi^2}\frac{1}{r_*^2}-\frac{1}{4\pi^2}
\frac{2(x^2{x'}^2+y^2{y'}^2)/r_*^2-r_*'^2}{r_*^4}+O({r_*'}^4/{r_*}^6)\,.
\end{eqnarray}
Hence, if $\theta=\tan^{-1}x/y=\pi/4$,
\begin{equation}
\left.\frac{1}{(2\pi)^2}\int_0^{2\pi}d\varphi_1'\int_0^{2\pi}d\varphi_2'
\,G(\mbox{\boldmath
$r$}_*,\mbox{\boldmath
$r$}_*')\right|_{\theta=\pi/4}=-\frac{1}{4\pi^2}\frac{1}{r_*^2}
+O({r_*'}^4/{r_*}^6)\,.
\end{equation}
Therefore, under the assumption that the higher multipole moments
are relatively small, we take an approximation
\begin{equation}
\mu_*\approx -\frac{1}{4\pi^2}\frac{1}{\bar{R}_*^2}\int d^{4}
\mbox{\boldmath $r$}_*\,\rho(\mbox{\boldmath $r$}_*) =
-\frac{1}{4\pi^2}\frac{1}{\bar{R}_*^2}\frac{4\pi
Gm^3N}{\Lambda}\,,
\end{equation}
where $\bar{R}_*$ denotes the distance between the origin and the boson
star surface in the direction of $\theta=\pi/4$.%
\footnote{For four-dimensional boson stars, the
value of $\mu_*$ obtained through the surface potential at
$\cos\theta=1/\sqrt{3}$ agrees with the value obtained by 
integration within a deviation of at most ten percent in the
region shown in Fig.~\ref{figen2}.}

%%%%%%%%%%%%%%%%%%%%%%%%%%%
%\begin{wrapfigure}{r}{5cm}
\begin{figure}[ht]
\centering
\includegraphics[height=5cm]
{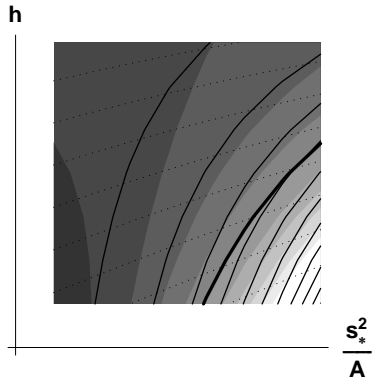}
\caption{%
The normalized Newtonian energy $E/N^2$ for the five-dimensional boson
star shown by gray shades. On the thick line, the energy
vanishes ($E=0$). The dotted lines indicate contours of equal radii, while
the solid lines indicate contours of equal
$s_*^2/N$.}
\label{fig5en2}
\end{figure}
%\end{wrapfigure}
%%%%%%%%%%%%%%%%%%%%%%%%%%%

Table~\ref{taen2} lists the values of $({\Lambda}/(4\pi Gm^4))E/N^2$
for the five-dimensional boson star
as a function of $h$ and $s_*^2/A$.
If $({\Lambda}/(4\pi Gm^4))E/N^2$ is negative, the solution is
considered to be energetically stable.

%%%%%%%%%%%%%%%%%%%%%%%%%%%
\begin{table}[htb]
\begin{center}
{\scriptsize
\begin{tabular}{|c|c|cccccccc|}\hline
&&&&&\multicolumn{2}{c}{$s_*^2/A$}&&&\\ \hline
& & 0.01 & 0.02 & 0.03 & 0.04 & 0.05 & 0.06 & 0.07 & 0.08\\ \hline
&0.1 &-0.00139555 & -0.00117091 & -0.000879967 & 
    -0.00048938 & 0.0000448016 & 0.000796971 & 
   0.00185744 & 0.00342503 \cr
&0.2 &-0.00133381 & 
    -0.00116085 & -0.000939441 & -0.000646538 & 
    -0.000251982 & 0.000288229 & 0.00104247 & 
   0.00211638 \cr 
&0.3 &-0.00127291 & -0.00114182 & 
    -0.000976895 & -0.000763081 & -0.000481243 & 
    -0.000105183 & 0.000407864 & 0.0011074 \cr
$h$ &0.4 &-0.0012138 & -0.001116 & -0.000996233 & 
    -0.000844575 & -0.00064993 & -0.000398235 & 
    -0.0000668219 & 0.000374118 \cr
&0.5 &-0.0011569 & 
    -0.00108564 & -0.00100074 & -0.000896475 & 
    -0.000766664 & -0.000603806 & -0.00039788 & 
    -0.000134626 \cr
&0.6 &-0.0011025 & -0.00105179 & 
    -0.000993517 & -0.000924326 & -0.00084106 & 
    -0.000740175 & -0.000617337 & -0.000466712 \cr
&0.7 &-0.00105065 & -0.00101573 & -0.000977242 & 
    -0.000933241 & -0.000882255 & -0.00082267 & 
    -0.000752846 & -0.000670658 \\ \hline
\end{tabular}
}
\caption{The values of $({\Lambda}/(4\pi Gm^4))E/N^2$ for the
five-dimensional boson star.%G4
}
\label{taen25}
\end{center}
\end{table}
%%%%%%%%%%%%%%%%%%%%%%%%%%%

Figure~\ref{fig5en2} shows the normalized binding energy.
The horizontal axis represents $s_*^2/A=\{0.01,0.08\}$,
while the vertical axis represents $h=\{0.1,0.7\}$.
A darker region indicates a lower energy.
Each contour shows the line on which the value
$\frac{{\Lambda}}{4\pi G m^3}\frac{s_*^2}{N}=\frac{s^2}{4\pi
G m^3 N}$ is constant.

For a small $s^2/N$, a lower $h$ yields a lower $E/N^2$. Thus the
 radius $\bar{R}_*$ of the boson star is not significantly changed from
the spherical case for a small
$s^2/N$. For a large $s^2/N$, the Newtonian binding energy becomes lower
at a finite $h$. This qualitative behavior is much similar to the case for
the boson star in four dimensions.

%%%%%%%%%%%%%%%%%%%%%%%%%%%
%\begin{wrapfigure}{r}{5cm}
\begin{figure}[ht]
\centering
\includegraphics[height=4cm]
{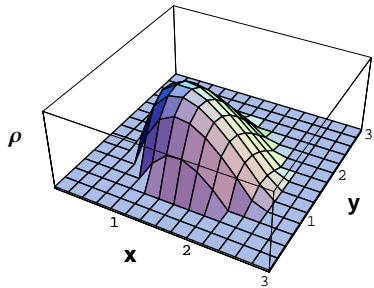}
\caption{%
A profile of the Newtonian rotating boson star
with the parameters $h=0.4$ and $s_*^2/A=0.06$.
}
\label{fig5sh}
\end{figure}
%\end{wrapfigure}
%%%%%%%%%%%%%%%%%%%%%%%%%%%

Figure~\ref{fig5sh} shows a typical profile of the five-dimensional
rotating boson star.

Brihaye and Hartmann reported \cite{BH} that the minimal boson star (in
the case with no self-coupling of the scalar field) in five dimensions is
energetically unstable.
Although they attributed this instability to the power of the long-range
tail of the gravitational potential, it is obvious that the stability
depends on the balance between the self-interaction of the scalar field
and the gravity. Therefore, the Newtonian approach in the higher
dimensions will be a useful as well as significant method to study
boson stars in models with general scalar potentials.

%%%%%%%%%%%%%%%%%%%%%%%%%%%%%%%%%%%%%%%%%%%%%%%%%%%%%%%%%%%%%%%%%
%%%%%%%%%%%%%%%%%%%%%%%%%%%%%%%%%%%%%%%%%%%%%%%%%%%%%%%%%%%%%%%%%
\section{Summary and prospects}
\label{sec7}
%%%%%%%%%%%%%%%%%%%%%%%%%%%%%%%%%%%%%%%%%%%%%%%%%%%%%%%%%%%%%%%%%
%%%%%%%%%%%%%%%%%%%%%%%%%%%%%%%%%%%%%%%%%%%%%%%%%%%%%%%%%%%%%%%%%

In this paper, Newtonian boson stars in the large coupling limit have been
studied by means of various approximation methods. In summary, we found
the qualitative behavior of the boson star parameters: 1) The Newtonian
binding energy is determined by the particle number and the value of the
gravitational potential, in addition to the energy from the rotation. 2)
With a certain rapid rotation, the radius of the boson star becomes
larger than that of the spherical boson star, if the particle number is
fixed. 3) For a slowly rotating boson star, the radius is not
significantly changed from that of the spherical one. 4) For a rotating
four-dimensional boson star, the ratio of the quadrupole moment to the
square of the angular momentum decreases when the angular momentum
increases if its radius is unchanged from that of the spherical one. 5)
The ratio of the quadrupole moment to the square of the angular momentum
increases when the radius of the boson star increases.

We used {\tt Mathematica} 4.2 \cite{wolfram} on a personal computer
for numerical calculations. Of course, elaborate calculation on a
large system would yield better results in terms of quality as well as 
quantity. In the present paper, since we used the minimal number
of parameters to describe the deformation of the rotating boson star,
we limit ourselves to a qualitative conclusion. The most important
subject to be examined is the introduction of more parameters which
determine the shape of the boson star in detail. This also depends on the
computational environment.

The boson stars with finite self-coupling can also be approximated
analytically by connecting the exponential tail in the asymptotic region
of the boson star \cite{GA}. It is interesting to study the stability of
the boson star in higher dimensions with analytical approximations in the
Newtonian limit.

We wish to solve the configuration of the binary of the boson
stars in a similar methods shown in this paper. The Newtonian treatment of
the system would shed light on a possible relation among the
physical quantities of the binary boson stars and provide an initial
state for the dynamical calculation with general relativity or other
theories of gravity. 
We also suppose that a rotating boson stars with a point mass at its
center resembles a rotating black hole with scalar hair
\cite{HRR,HR1,HR2}. The study of such systems with the Newtonian
approximation is interesting and may reveal an essential nature of
gravity.

Finally, we anticipate that the Newtonian approach to the boson star would
be valid for a multi-scalar system with many $U(1)$ charges, such as the
`multi-state boson star' \cite{BBAP}, and the boson-fermion system
\cite{HLM1,HLM2,HLM3,J2}, in order to extract its inherent
characteristics in an efficient manner.

%\acknowledgments
%%%%%%%%%%%%%%%%%%%%%%%%%%%%%%%%%%%%%%%%%%%%%%%%%%%%%%%%%%%%%%%%%%%%%%%%%%%
%Acknowledgements
%%%%%%%%%%%%%%%%%%%%%%%%%%%%%%%%%%%%%%%%%%%%%%%%%%%%%%%%%%%%%%%%%%%%%%%%%%%
\begin{acknowledgments}
We thank Prof. Carlos Herdeiro for valuable information on
their work on rotating boson stars and hairy black holes
\cite{HRR,HR1,HR2}.
%the organizers of JGRG21, where our
%partial result %({\tt [arXiv:10mm.xxxx]}) 
%was presented. %for elucidating comments.
%This study is supported in part by the Grant-in-Aid of Nikaido Research 
%Fund.
\end{acknowledgments}

%%%%%%%%%%%%%%%%%%%%%%%%%%%%%%%%%%%%%%%%%%%%%%%%%%%%%%%%%%%%%%%%%%%%%%%%%%%
%%%%%%%%%%%%%%%%%%%%%%%%%%%%%%%%%%%%%%%%%%%%%%%%%%%%%%%%%%%%%%%%%%%%%%%%%%%
%%%%%%%%%%%%%%%%%%%%%%%%%%%%%%%%%%%%%%%%%%%%%%%%%%%%%%%%%%%%%%%%%%%%%%%%%%%
%%%%%%%%%%%%%%%%%%%%%%%%%%%%%%%%%%%%%%%%%%%%%%%%%%%%%%%%%%%%%%%%%%%%%%%%%%%

%%%%%%%%%%%%%%%%%%%%%%%%%%%%%%%%%%%%%%%%%%%%%%%%%%%%%%%%%%%%%%%%%%%%%%%%%%%

%%%%%%%%%%%%%%%%%%%%%%%%%%%%%%%%%%%%%%%%%%%%%%%%%%%%%%%%%%%%%%%%%%%%%%%%%%%
%thebibliography
%%%%%%%%%%%%%%%%%%%%%%%%%%%%%%%%%%%%%%%%%%%%%%%%%%%%%%%%%%%%%%%%%%%%%%%%%%%
%\bibliographystyle{apsrev}
\bibliographystyle{apsrev4-1}
%\bibliography{}

\begin{thebibliography}{99}

\bibitem{BH}
Y.~Brihaye and B.~Hartmann, 
%Minimal boson stars in 5 dimensions
Class. Quant. Grav. {\bf 33} (2016) 065002.
%arXiv:1509.04534.

\bibitem{HRS}
B.~Hartmann, J. Riedel and R. Suciu,
%Gauss--Bonnet boson stars
Phys Lett. {\bf B726} (2013) 906.%--912.

\bibitem{HKKL}
B.~Hartmann, B.~Kleihaus, J.~Kunz and M.~List, 
%Rotating boson stars in 5 dimensions
Phys. Rev. {\bf D82} (2010) 084022.
%arXiv:1008.3137.

\bibitem{Jetzer} P.~Jetzer, 
%Boson stars
Phys. Rep. {\bf 220} (1992) 163.%--227

\bibitem{LM} A.~R.~Liddle and M.~S.~Madsen, 
%The Structure and formation of boson stars
Int. J. Mod. Phys. {\bf D1} (1992) 101.%--144.

\bibitem{SM} F.~E.~Schunck and E.~W.~Mielke, 
Class. Quant. Grav. {\bf 20} (2003) R301.
%arXiv:0801.0307 [astro-ph].

\bibitem{R1SL1} F.~E.~Schunck and A.~R.~Liddle, 
%The gravitational redshift of boson stars
Phys. Lett. {\bf B404} (1997) 25.%--32.

\bibitem{R1SL2} F.~E.~Schunck and A.~R.~Liddle, 
``Boson stars in the centre of galaxies?'' in
{\it ``Black Holes: Theory and Observation'',
Proceedings of the Bad Honnef Workshop},
F.~W.~Hehl, C.~Kiefer and R.~J.~K.~Metzler eds. (Springer-Verlag, Berlin,
1998), pp. 285--288, arXiv:0811.3764 [astro-ph].

\bibitem{TCL} D.~F.~Torres, S.~Capozziello and G.~Lambiase,
%Supermassive boson star at the galactic center?
Phys. Rev. {\bf D62} (2000) 104012.

\bibitem{ST} F.~E.~Schunck and D.~F.~Torres,
%Boson stars with generic self-interactions
Int. J. Mod. Phys. {\bf D9} (2000) 601.
%arXiv:gr-qc/9911038.

\bibitem{MA}
T.~Matos and L.~A.~Ure\~na-L\'opez, 
%Flat rotation curves in scalar field galaxy halos
Gen. Rel. Grav. {\bf 39} (2007) 1279.%-1286.

\bibitem{R1S}
F.~E.~Schunck, {\it ``Selbstgravitierende bosonische Materie''}
PhD-thesis, University of Cologne; 1996 (Cuvillier Press: G\"ottingen).

\bibitem{R1SM1}
F.~E.~Schunck and  E.~W.~Mielke, ``Rotating
boson stars'' in {\it ``Relativity
and Scientific Computing: Computer Algebra, Numerics, Visualization'',
Proceedings of the Bad Honnef Workshop},
F.~W.~Hehl, R.~A.~Puntigam and H.~Ruder eds. (Springer-Verlag, Berlin,
1996), pp. 138--151.

\bibitem{R1SM2}
F.~E.~Schunck and  E.~W.~Mielke,
%Rotating boson star as an effective mass torus in general relativity
Phys. Lett. {\bf A249} (1998) 389.%--394.

\bibitem{Ryan}
F.~D.~Ryan,
%Spinning boson stars with large selfinteraction
Phys. Rev. {\bf D55} (1997) 6081.%--6091.

\bibitem{HRR}
C.~A.~R.~Herdeiro, E.~Radu and H.~R\'unarsson,
%Kerr black holes with self-interaction scalar hair
Phys. Rev. {\bf D92} (2015) 084059.

\bibitem{colpi} M.~Colpi, S.~L.~Shapiro and I.~Wasserman, 
%Boson Stars: Gravitational Equilibria of Selfinteracting Scalar Fields
Phys. Rev. Lett. {\bf 57} (1986) 2485.%--2488.

\bibitem{wolfram}
Wolfram, http://www.wolfram.com/.

\bibitem{GL}
R. Gregory and R. Laflamme,
Phys. Rev. Lett. {\bf 70} (1993) 2837.%--2840.

\bibitem{GA}
F. S. Guzm\'an and L. A. Ure\~na-L\'opez,
%Newtonian collapse of scalar field dark matter
Phys. Rev. {\bf D68} (2003) 024023.

\bibitem{HR1}
C.~A.~R.~Herdeiro and E.~Radu,
%Kerr black holes with scalar hair
Phys. Rev. Lett. {\bf 112} (2014) 221101.

\bibitem{HR2}
C.~A.~R.~Herdeiro and E.~Radu,
%Construction and physical properties of Kerr black holes 
%with scalar hair
Class. Quant. Grav. {\bf 32} (2015) 144001.

\bibitem{BBAP} A.~Bernal, J.~Barranco, D.~Alic and C.~Palenzuela,
% multi-state boson stars
Phys. Rev. {\bf D81} (2010) 044031.
% arXiv:0908.2435 [gr-qc].

\bibitem{HLM1}
A. B. Henriques, A. R. Liddle and R. G. Moorhouse, 
%Combined Boson - Fermion Stars
Phys. Lett. {\bf B233} (1989) 99.

\bibitem{HLM2}
A. B. Henriques, A. R. Liddle and R. G. Moorhouse,
%Combined Boson - Fermion Stars: Configurations and Stability
Nucl. Phys. {\bf B337} (1990) 737. 

\bibitem{HLM3}
A. B. Henriques, A. R. Liddle and R. G. Moorhouse,
%Stability of boson-fermion stars
Phys. Lett. {\bf B251} (1989) 511.%--516 

\bibitem{J2}
Ph. Jetzer, 
%Stability of Combined Boson-Fermion Stars
Phys. Lett. {\bf B243} (1990) 36.%--40.






\end{thebibliography}

%%%%%%%%%%%%%%%%%%%%%%%%%%%%%%%%%%%%%%%%%%%%%
%%%%%%%%%%%%%%%%%%%%%%%%%%%%%%%%%%%%%%%%%%%%%
%%%%%%%%%%%%%%%%%%%%%%%%%%%%%%%%%%%%%%%%%%%%%
%%%%%%%%%%%%%%%%%%%%%%%%%%%%%%%%%%%%%%%%%%%%%
%%%%%%%%%%%%%%%%%%%%%%%%%%%%%%%%%%%%%%%%%%%%%
%%%%%%%%%%%%%%%%%%%%%%%%%%%%%%%%%%%%%%%%%%%%%
%%%%%%%%%%%%%%%%%%%%%%%%%%%%%%%%%%%%%%%%%%%%%

%%%%%%%%%%%%%%%%%%%%%%%%%%%%%%%%%%%%%%%%%%%%%%%%%%%%%%%%%%%%%%%%%%%%%%%%%%%
%%%%%%%%%%%%%%%%%%%%%%%%%%%%%%%%%%%%%%%%%%%%%%%%%%%%%%%%%%%%%%%%%%%%%%%%%%%
%%%%%%%%%%%%%%%%%%%%%%%%%%%%%%%%%%%%%%%%%%%%%%%%%%%%%%%%%%%%%%%%%%%%%%%%%%%
%%%%%%%%%%%%%%%%%%%%%%%%%%%%%%%%%%%%%%%%%%%%%%%%%%%%%%%%%%%%%%%%%%%%%%%%%%%
%%%%%%%%%%%%%%%%%%%%%%%%%%%%%%%%%%%%%%%%%%%%%%%%%%%%%%%%%%%%%%%%%%%%%%%%%%%

%%%%%%%%%%%%%%%%%%%%%%%%%%%%%%%%%%%%%%%%%%%%%%%%%%%%%%%%%%%%%%%%%%%%%%%%%%%
%%%%%%%%%%%%%%%%%%%%%%%%%%%%%%%%%%%%%%%%%%%%%%%%%%%%%%%%%%%%%%%%%%%%%%%%%%%
%%%%%%%%%%%%%%%%%%%%%%%%%%%%%%%%%%%%%%%%%%%%%%%%%%%%%%%%%%%%%%%%%%%%%%%%%%%
%%%%%%%%%%%%%%%%%%%%%%%%%%%%%%%%%%%%%%%%%%%%%%%%%%%%%%%%%%%%%%%%%%%%%%%%%%%
%%%%%%%%%%%%%%%%%%%%%%%%%%%%%%%%%%%%%%%%%%%%%%%%%%%%%%%%%%%%%%%%%%%%%%%%%%%

\end{document}